# Extrinsic and Intrinsic Anomalous Metallic States in Transition Metal Dichalcogenide Ising Superconductors

Ying Xing,[∥,†,‡] Pu Yang,[∥,Δ,‡] Jun Ge,[†,‡] Jiaojie Yan,[†,‡] Jiawei Luo,[†] Haoran Ji,[†] Zeyan Yang,[∥] Yongjie Li,[∥] Zijia Wang,[∥] Yanzhao Liu,[†] Feng Yang,[∥] Ping Qiu,[∥] Chuanying Xi,[◊] Mingliang Tian,[◊] Yi Liu,[#,*] Xi Lin,[†,§,⊥,*] Jian Wang,[†,§,⊥,*]

[∥]State Key Laboratory of Heavy Oil Processing, College of New Energy and Materials, China University of Petroleum, Beijing 102249, China.

[†]International Center for Quantum Materials, School of Physics, Peking University, Beijing 100871, China.

[Δ]College of Chemistry, Beijing Normal University, Beijing 100875, China.

[◊]High Magnetic Field Laboratory, Chinese Academy of Sciences, Hefei 230031, China.

[#]Department of Physics, Renmin University of China, Beijing 100872, China.

[§]CAS Center for Excellence in Topological Quantum Computation, University of Chinese Academy of Sciences, Beijing 100190, China.

[⊥]Beijing Academy of Quantum Information Sciences, Beijing 100193, China.

[†] Corresponding author. E-mail: jianwangphysics@pku.edu.cn (J.W.); xilin@pku.edu.cn (X.L.); yiliu@ruc.edu.cn (Yi L.)

ABSTRACT

The metallic ground state in two-dimensional (2D) superconductors has attracted much attention but is still under intense scrutiny. Especially, the measurements in ultralow temperature region are challenging for 2D superconductors due to the sensitivity to external perturbations. In this work, the resistance saturation induced by external noise, named as "extrinsic anomalous metallic state", is observed in 2D transition metal dichalcogenide (TMD) superconductor $4Ha$-TaSe$_2$ nanodevices. However, with further decreasing temperature, credible evidence of intrinsic anomalous metallic state is obtained by adequately filtering external radiation. Our work indicates that at ultralow temperatures the anomalous metallic state can be experimentally revealed as the quantum ground state in 2D crystalline TMD superconductors. Besides, Ising superconductivity revealed by ultrahigh in-plane critical field ($B_{c2//}$) going beyond the Pauli paramagnetic limit ($B_p$) is detected in $4Ha$-TaSe$_2$, from one-unit-cell device to bulk situation, which might be due to the weak coupling between the TaSe$_2$ sub-monolayers.

TEXT

2D crystalline superconductors[1], distinct from amorphous and granular thin films[2, 3], usually exhibit a square resistance far below the quantum resistance $R = h/(2e)^2 \approx 6450\ \Omega$ for Cooper pairs, with $h$ the Plank constant and $e$ the elementary charge. Among many exotic and popular phenomena, metallic ground state[4-8] and Ising superconductivity[9-11] are striking discoveries in 2D crystalline superconductors.

The quantum ground states for 2D bosonic systems should be either a superconducting state or an insulating state due to the conjugacy between the number and phase fluctuation of Cooper pairs[12]. However, the intermediate metallic state showing an anomalous temperature-independent resistance that far below the prediction of Drude formula when approaching absolute zero temperature, has been reported in a variety of 2D superconductors[3-8, 12-16]. In the study of ground state in 2D superconductors, the ultra-low temperature transport measurements are necessary but quite challenging. Because the high frequency noise from the instruments and external environmental effects usually affect the measurement results in ultralow temperature region, especially for the nanosystems showing fragile 2D superconductivity. A recent report[17] reveals that in crystalline $2H$-NbSe$_2$ nanodevices, the "metallic states" can be eliminated by adequately filtering external radiation. This work raises doubts about previous experimental observations on anomalous metallic state without appropriate filters. It is noteworthy that, in the molecular beam epitaxy grown macroscopic PdTe$_2$ films with robust 2D superconductivity, the anomalous metallic state is observed with high quality filters installed.[7] This motivates the further investigation on the universality of intrinsic ground states in crystalline 2D superconductors, especially for nanodevices showing relatively fragile superconductivity.

Recently, Ising superconductivity with large in-plane critical field $B_{c2//}$ (far beyond $B_p$) has been reported in gated $2H$-MoS$_2$[9, 11], sub-monolayer NbSe$_2$[10, 18], TaS$_2$[19], interface modulated ultrathin Pb films[20], few-layer $1T_d$-MoTe$_2$[21] and $3R$-TaSe$_2$[22] in the absence of in-plane inversion symmetry. In the 2D limit, the broken in-plane inversion symmetry can give rise to the Zeeman-type spin-orbit interaction (SOI), which pins the electron spins to the out-of-plane direction, and then reduces the spin pair-breaking effect of the in-plane field. With increasing thickness, the interlayer coupling will break such spin-momentum locking and induce the orbital effect, which can destroy the perfect

Ising pairing and strongly reduce the $B_{c2//}$.

Layered TMDs, with weak van der Waals interlayer bonding between adjacent layers, provide an ideal material platform to realize crystalline 2D superconductors by mechanical exfoliation etc. In TMDs, the special stacking styles of octahedral or trigonal prismatic TMD blocks allow many polytypes to form. Among them, 4*Ha* phase[23] (schematic crystal structure is shown in Supporting Information Fig. S1(a)) is a special case with broken in-plane inversion symmetry from individual layers to bulk situation. The number 4 in 4*Ha* indicates the number of individual layers in one-unit-cell, and *H* represents the hexagonal structure.

In this work, we present systematic electrical transport studies on crystalline non-centrosymmetric superconductor 4*Ha*-TaSe$_2$, from bulk to 2.8 nm (~ 1-unit-cell, 1-UC). The external radiation induced resistance saturation (named as extrinsic anomalous metallic state) at low temperatures is observed in 4*Ha*-TaSe$_2$ nanodevices. By sufficiently filtering the external perturbations in the electrical transport measurements, the extrinsic anomalous metallic state becomes superconducting state showing zero resistance within the measurement resolution. However, at lower temperatures and higher magnetic fields, we provide the evidence of intrinsic anomalous metallic state in TaSe$_2$ nanodevices with high quality filters. Furthermore, for 2.8 nm 4*Ha*-TaSe$_2$, the measured data at ultralow temperatures and high magnetic fields point to the in-plane $B_{c2//}$ ~ 3.9 times of its $B_p$, demonstrating the Ising superconductivity. Surprisingly, the $B_{c2//}$ enhancement persists from ~1-UC to the bulk, which may originate from the weak coupling between individual layers in 4*Ha* phase.

We synthesized 4*Ha*-TaSe$_2$ single crystals by the chemical vapor transport method and characterized them using X-ray diffraction (XRD), transmission electron microscopy (TEM), Raman spectroscopy and energy dispersive spectrometer (EDS). An optical image of a typical TaSe$_2$ single crystal is shown in the inset of Supporting Information Fig. S1(b). Figure 1(a) shows the room temperature XRD pattern from a typical TaSe$_2$ single crystal oriented with the scattering vector perpendicular to the (001) plane. The lattice constant along the *c*-axis is calculated to be 25.501 Å. The single crystalline feature of TaSe$_2$ is further revealed by TEM measured at room temperature (Fig. 1(a), inset). The EDS mapping confirms the chemical fingerprint of sample, with Ta and Se signals emerging only inside the TaSe$_2$ flake (Fig. 1(b)). Characteristic Raman signals of TaSe$_2$ flakes at room temperature with thickness various from 52.0 nm to 4.2 nm are shown in Fig. 1(c). Three TaSe$_2$ characteristic peaks of E$_{1g}$, E$_{2g}$, A$_{1g}$ are detected, corresponding to intralayer (E$_{1g}$, E$_{2g}$) and interlayer (A$_{1g}$) vibration modes respectively.[24, 25]

In our experiments, we fabricated few-layer TaSe$_2$ devices by mechanical exfoliation of single crystals followed by direct transfer onto 300-nm-thick SiO$_2$/Si substrates. The standard e-beam lithography followed by e-beam evaporation was used to fabricate electrodes (see more details in Supporting Information). Atomic force microscopy (AFM) measurements were performed to evaluate the morphology and thickness of 4$Ha$-TaSe$_2$ nanodevices (Fig. 1(d)).

The temperature dependence of the normalized four-point resistance $R(T)/R(300\ K)$ at zero magnetic field is shown in Fig. 2(a) from 0.5 K to 300 K. For bulk 4$Ha$-TaSe$_2$, a resistance anomaly representing charge density wave transition is found around 115 K. While for nanodevices with thickness smaller than 8.6 nm, no obvious anomaly can be detected in $R(T)/R(300\ K)$ curves. All devices show metallic behavior with $R\propto T$ in high temperature region, reminiscent of strange metal state in high $T_c$ superconductors[26] and then approach a constant value before going to the superconducting state. The inset of Fig. 2(a) is an optical image of a measured device (8.6 nm). The residual resistance ratio defined as the ratio of room temperature resistance and the normal state resistance right above the superconducting transition ~ $R(300\ K)/R(5\ K)$ varies from 8.1 (bulk) to 2.5 (2.8 nm device). The square resistance (at 5 K) is estimated to be 930 Ω for 2.8 nm device, 30 ~ 500 Ω for thicker devices. These values are much smaller than quantum resistance $h/(2e)^2$ for Cooper pairs, the critical resistance of a disorder induced superconductor-insulator transition. This implies our samples are in the low disorder region. Superconductivity is observed for all samples from bulk material to 2.8 nm nanodevice. The onset ($T_c^{onset}$) and zero-resistance ($T_c^{zero}$) superconducting critical transition temperatures of the bulk 4$Ha$-TaSe$_2$ (Supporting Information Fig. S2(a)) are 2.92 K and 2.66 K respectively, with a narrow transition width of 0.26 K. Here, $T_c^{zero}$ is identified as resistance falls below the instrument resolution limit. Both $T_c^{onset}$ and $T_c^{zero}$ in few-layer TaSe$_2$ devices are slightly reduced than those of the bulk sample.

The 2D nature of the superconducting TaSe$_2$ nanodevices is confirmed by the experiments with tilted magnetic field. Figure 2(b) shows the angular dependence of the upper critical field $B_{c2}(\theta)$ ($\theta$ represents the angle between the $c$ axis of TaSe$_2$ and applied field direction) in a 6.0 nm TaSe$_2$ nanodevice at 1.45 K. In this paper, we define the critical field $B_{c2}$ at a finite temperature as the field corresponding to 50% of normal resistance ($R_N$). A cusp-like peak is clearly resolved at $\theta = 90°$ and qualitatively distinct from the 3D anisotropic mass model but can be well described by the 2D Tinkham model, which yields $B_{c2//} \sim 8.27$ T, $B_{c2\perp} \sim 0.15$ T[27]. Notably, the anisotropic parameter $\gamma = B_{c2//}/B_{c2\perp}$ is about 55.13 for 6.0 nm TaSe$_2$ device but only 6 for bulk TaSe$_2$ (Supporting Information Fig. S2).

Normally, low dimensional superconductivity is easy to be affected or even destroyed by external perturbations, especially in ultralow temperature region. To investigate the ground state of 4$Ha$-TaSe$_2$ devices, the controlled experiments on filtered and unfiltered measurements in a dilution refrigerator (DR) were carried out. The data in Fig. 2(c)-(d) and Fig. 3 were obtained from two 4$Ha$-TaSe$_2$ devices possessing similar thickness (5.0 nm and 7.0 nm), normal square resistance and superconducting properties. Figure 2(c) displays two sets of $R_S(T)$ curves of 5.0 nm TaSe$_2$ device ($T_c^{onset} \sim 2.02$ K, $T_c^{zero} \sim 1.72$ K) under out-of-plane magnetic field with $T$ down to 0.5 K (Leiden CF450), with and without silver-epoxy filters installed in each lead of samples. For both curves with same magnetic field (plotted by same color), the bottom ones are filtered data and the top ones are unfiltered. Figure 2(d) shows the $R$ versus $T^{-1}$ data corresponding to (c).

The silver-epoxy filters exhibit significant impact on low temperature electrical transport measurements of TaSe$_2$ devices. For instance, at 0.45 T, the largest $\Delta R$ between filtered and unfiltered measurements at 0.56 K is $\sim 44$ $\Omega$. For unfiltered measurements, the square resistance under small magnetic field initially decreases exponentially with an approximate thermal activated flux flow behavior $R(T) \propto \exp(-U(B)/k_B T)$,[5, 28, 29] and then saturates in low temperature region (Supporting Information Fig. S3). Here the excitation gap $U(B)$ is the free energy barriers coming from the pinning effect of physical defects, such as inhomogeneity, strains, dislocations. $U(B)$ decreases with increasing magnetic field due to the Lorentz force. However, these $R$ saturations are restored when home-made silver-epoxy filters are installed into the measurement setup (Supporting Information Fig. S4(a)). Especially for magnetic field at 0.10 T, 0.20 T and 0.30 T, $R$ returns to zero within the measurement resolution directly from finite platform. Here, the resistance saturation induced by external high frequency noise is named as extrinsic anomalous metallic state. These results are similar to previous report on 2$H$-NbSe$_2$ nanodevices by Tamir $et$ $al$.[17]

Furthermore, the intrinsic anomalous metallic state is characterized by the resistance saturation at ultralow temperatures, which is the intrinsic property of 2D superconducting system rather than a result of extrinsic factors such as high frequency noise, large excitation current etc. To further figure out whether the intrinsic anomalous metallic state exists in crystalline TMD devices at ultralow temperatures, we measured 7.0-nm-thick TaSe$_2$ device ($T_c^{onset}$~2.05 K, $T_c^{zero}$~1.90 K) down to 0.055 K by another DR (Leiden CF-CS81-600) equipped with high quality silver-epoxy filters at low temperature stage and room temperature resistance-capacitor (RC) filters (Supporting Information Fig. S4(b)). The combination of the silver-epoxy filter and RC filter in series reaches noise floor in the GHz range and reduces to ~0.0001% of the initial noise power in the MHz range[30]. Figure 3(a) shows the $R_S(T)$ curves under different out-of-plane magnetic fields from 0.704 T to 1.254 T. For magnetic field larger than 1.254 T, the sample enters normal state for all measured temperatures. The corresponding $R_S(B)$ curves are displayed in Supporting Information Fig. S5. For the field below 1.254 T, a resistance drop is observed on cooling from the normal state. Below $T_c$, the device exhibits thermally activated flux flow behavior (linear slope in the Arrhenius plot, Fig. 3(b)). As the temperature further decreasing till to the lowest temperature ~0.055 K, the resistance deviates from the linear behavior and saturates to a finite value (dependent on specific magnetic field > 0.704 T). The inset of Fig. 3(c) shows the linear voltage-current behavior below 50 nA at 0.055 K and 0.854 T. The excitation current used in Fig. 3(a) is 10~20 nA. This guarantees the observed resistance saturation in the ultralow temperature region is reliable. Therefore, the $R$ saturation reenters in lower temperature and higher magnetic field region, excluding the interference of external noise in TaSe$_2$ device, demonstrating the intrinsic anomalous metallic state in crystalline 4$Ha$-TaSe$_2$ device. The above observations are also confirmed in another 4.4 nm TaSe$_2$ device (Supporting Information Fig. S7). Our findings manifest that although previous observations of $R$ saturation in the absence of filters are induced by external noise, at lower temperatures and higher magnetic fields the intrinsic anomalous metallic state still exists in two-dimensional crystalline superconductors.

Next, we determine the thermal activation energy $U$ [5, 28, 29] from the slope of linear portion in the Arrhenius plot (black solid line in Fig. 3(b)) at different magnetic fields, and the results are summarized in Fig. 3(c). A fit to $U(B)=U_0\ln(B_0/B)$ yields $U_0 = 6.29$ K and $B_0 = 0.91$ T (approximate to the experimental value of $B_{c2}^{\perp}$ ~ 0.93 T) for 7.0 nm TaSe$_2$ device. With decreasing temperature, the thermal activated vortex motion (thermal creep) is replaced by the quantum tunneling of vortices (quantum creep)[31], one possible description of the metallic state in 4H-TaSe$_2$ devices. In this model, the vortices hop between the local potential minima through the quantum tunneling, and the resistance obeys a general form in the limit of the strong dissipation:

$$R_S \sim \frac{\hbar}{4e^2}\frac{\kappa}{1-\kappa}, \quad \kappa \sim \exp\left[C\frac{\hbar}{e^2}\frac{1}{R_N}\left(\frac{B-B_{c2}^{\perp}}{B}\right)\right] \tag{1}$$

where $C$ is a dimensionless constant of order unity, $R_N$ is the square resistance of the normal state and $B_{c2}^{\perp}$ is the out-of-plane upper critical field. Figure 3(d) displays $R(B_{\perp})$ curve of 7.0 nm TaSe$_2$ device at 0.055 K which can be well fitted by Eq.(1) below 0.89 T. Above 0.89 T, $R$ is well described by a $B$-linear dependence, suggesting pinning-free vortex flow. The crossover from the quantum creep to the flow motion occurs at 0.89 T, close to where the $U(B)$ for the thermal creep approaches zero (0.91 T, Fig. 3(c)), implying that the pinning or the elastic potential effectively disappears at high magnetic fields.

Figure 4 summarizes the phase diagram of the TaSe$_2$ nanodevices in perpendicular magnetic field. The boundary of zero resistance region at low temperatures and low magnetic fields (green region) is defined by TaSe$_2$ devices with high quality filters. Below $B/B_{c2}$ ~0.55, with increasing temperature, the TaSe$_2$ nanodevice in unfiltered measurement evolves from a resistance saturation state induced by external noise, i.e. "extrinsic anomalous metallic state" (white transparent region), to the thermal creep region, and followed by a transition to the normal state at $B_{c2}$. The intrinsic anomalous metallic state detected by the combined high quality filters locates at the ultralow temperature and relatively high magnetic field region (see blue region), corresponding to the quantum ground state. Finally, we would like to clarify the correlation between this work and ref 17. The lowest temperature for crystalline NbSe$_2$ nanodevices in ref 17 is 0.250 K and no signature of resistance saturation is found. In our case, we present experimental results of TaSe$_2$ nanodevices at lower temperatures down to 0.055 K and demonstrate the intrinsic anomalous metallic state in the TaSe$_2$ nanodevices showing relatively fragile superconductivity. Our work for the first time reveals that although the extrinsic metallic state induced by the external high frequency noise can be detected in 2D crystalline TMD superconductors, the intrinsic metallic state as a ground state still exists at lower temperatures.

Systematic magnetoresistance of 4$Ha$-TaSe$_2$ from bulk to 2.8 nm (close to 1-UC) under both out-of-plane ($B_\perp$, magnetic field applied perpendicular to the (001) surface) and in-plane field ($B_{//}$, magnetic field applied parallel to the (001) surface) were measured (see Fig. 5(a)-(c), Supporting Information Fig. S2(b)-(c) and Fig. S8). We summarize the $B_{c2}$-$T_c$ data in both directions in Fig. 5(d). For comparison, $B_{c2}$ is normalized by $B_p$ and the $T$ is normalized by $T_c$ (defined at 0.5 $R_N$) for each sample. The prominent large enhancement of $B_{c2//}$ is found in 2.8 nm TaSe$_2$ nanodevice: the measured $B_{c2//}$ at $T/T_c = 0.82$ is more than 3.9 times of the Pauli limit $B_p$. Following the previous works on other TMD Ising superconductors [9-11, 18, 19, 21], the $B_p$ [32, 33] here is estimated using standard Bardeen-Cooper-Schrieffer ratio as well as a g-factor of 2. The temperature dependence of in-plane $B_{c2//}$ of 2.8 nm TaSe$_2$ nanodevice is fitted by microscopic model for Ising pairing[20],

$$\ln\left(\frac{T}{T_c}\right) + \frac{\mu_B^2 B^2}{\widetilde{\beta_{SO}}^2 + \mu_B^2 B^2} \text{Re}\left[\psi\left(\frac{1}{2} + \frac{i\sqrt{\widetilde{\beta_{SO}}^2 + \mu_B^2 B^2}}{2\pi k_B T}\right) - \psi\left(\frac{1}{2}\right)\right] = 0 \qquad (2)$$

where $\psi(x)$ is the digamma function, $\widetilde{\beta_{SO}} = \beta_{SO}/[1 + \hbar/(2\pi k_B T_c \tau_0)]$ is the effective Zeeman-type SOI considering spin-independent scattering (3.43 meV for 2.8 nm nanodevice), and $\tau_0$ is mean free time. The microscopic model for Ising superconductivity, taking Zeeman-type SOI into account, matches the data well, further indicating that the depairing effect from the orbital magnetic field is strongly suppressed and the Ising pairing mechanism needs to be considered. Hence, $B_{c2//}$ is solely determined by the interaction between the magnetic field and the spin degree of freedom of the Cooper pairs.

For multi-layer 4$Ha$-TaSe$_2$ nanodevices, the interlayer coupling interaction may introduce the orbital effects and weaken the Ising pairing. As expected, the $B_{c2//}$ (filled symbols in Fig. 5(d)) decreases as the thickness increases, which yields smaller $\widetilde{\beta_{SO}}$ for thicker films ($\widetilde{\beta_{SO}}$ reduces to 1.63 meV in 5.0 nm TaSe$_2$ and 0.41 meV in 8.6 nm TaSe$_2$, see Supporting Information Fig. S9). Surprisingly, the $B_{c2//}$ of bulk 4$Ha$-TaSe$_2$ also exceeds its Pauli limit $B_{c2//}(0) \sim 2.0\ B_p$ (Fig. 5(b), Supporting Information Fig. S2(d)). While in 2$H$-TaSe$_2$ ($B_{c2//} \ll B_p$)[34] and other TMD superconductors like 2$H$-NbSe$_2$ ($B_{c2//} \gtrsim B_p$)[35], 2$H$-TaS$_2$ ($B_{c2//} \approx B_p$)[36], and pressured 1$T'$-MoTe$_2$ ($B_{c2//} \ll B_p$)[37], the bulk $B_{c2//}$ is approximate equal or less than $B_p$.

For few-layer samples, the orbital effect of the in-plane magnetic field is greatly suppressed due to the reduced dimensionality, and consequently $B_{c2//}$ is mainly determined by the interaction between the external magnetic field and the spins of the electrons. For thicker samples, as superconductivity is also influenced by orbital effects, the anisotropy and enhancement of $B_{c2//}$ are usually reduced. While for bulk $4Ha$-TaSe$_2$, the lattice constant along the $c$-axis obtained from our XRD data is 25.501 Å, larger than that of $2H$-TaSe$_2$ (12.696 Å for 1 UC and 25.392 Å for 4 Se-Ta-Se layers)[38], indicating larger distance between two adjacent layers and leading to weaker interlayer coupling accordingly. This may be the reason that the enhancement of $B_{c2//}$ sustains even in the $4Ha$-TaSe$_2$ bulk.

In conclusion, this work presents two novel quantum phenomena in thin TaSe$_2$ nanodevices: Ising superconductivity and anomalous metallic state. The large in-plane critical field as the characteristics of Ising superconductivity is detected not only in thin devices but also in bulk samples, which can be attributed to the weak interlayer coupling interaction in $4Ha$-TaSe$_2$. Moreover, under perpendicular magnetic field, the "anomalous metallic like behavior" in low temperature region is revealed to be caused by high-frequency noise. Interestingly, the lower temperature and higher magnetic field measurements with high quality filters manifest that the intrinsic anomalous metallic ground state still exists when approaching zero temperature. Thus, we demonstrate the presence of intrinsic anomalous metallic state even in the nanosystems with relatively fragile 2D superconductivity and indicate the universality of the anomalous metallic ground state in 2D crystalline superconductors. Our finding paves the way for further explorations and investigations on the nature of intrinsic anomalous metallic state in various 2D superconducting systems.

FIGURES

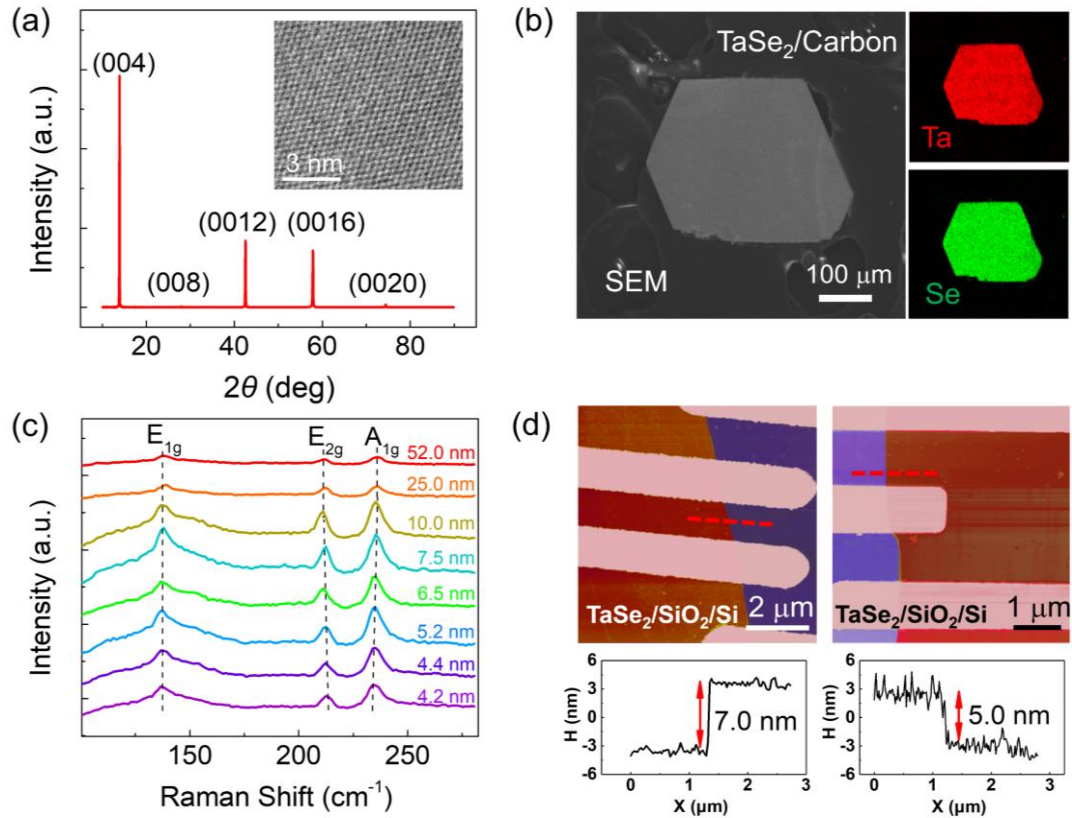

**Figure 1.** Sample characterizations of 4*Ha*-TaSe$_2$. (a) The XRD pattern of a TaSe$_2$ single crystal showing its 4*Ha* phase feature. Inset: atomic-resolution TEM image of TaSe$_2$ flake from the zone axis [001], revealing a hexagonal structure. (b) SEM image and EDS mapping images, the Ta and Se distribute evenly in TaSe$_2$ flake. (c) Thickness-dependence of Raman spectra for TaSe$_2$ on SiO$_2$/Si substrate with thickness varying from 4.2 nm to 52.0 nm. (d) AFM images (false-colored) and corresponding height profiles of two typical TaSe$_2$ nanodevices, showing thickness of 7.0 nm and 5.0 nm.

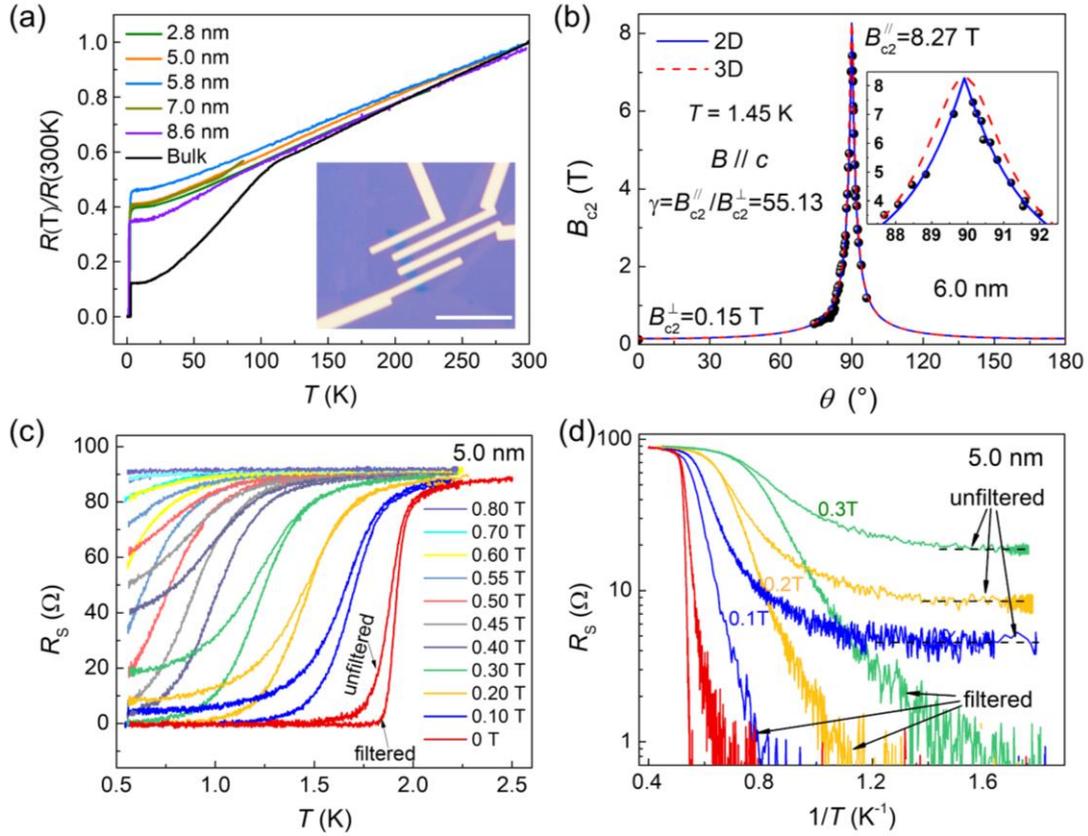

**Figure 2.** 2D superconductivity and extrinsic anomalous metallic state in 4$Ha$-TaSe$_2$ TaSe$_2$ nanodevices. (a) The temperature dependence of normalized resistance $R(T)/R(300\ \mathrm{K})$ of bulk TaSe$_2$ and few-layer nanodevices. The inset shows the optical image of 8.6-nm-thick TaSe$_2$ device. The scale bar represents 20 μm. (b) Angular dependence of the upper critical fields $B_{c2}(\theta)$ of 6.0-nm-thick TaSe$_2$ device ($\theta$ represents the angle between applied magnetic field and the perpendicular direction to the (001) surface of TaSe$_2$). The inset shows a close-up of the region around $\theta = 90°$. The blue solid line and the red dashed line correspond to the theoretical representations of $B_{c2}(\theta)$, using the 2D Tinkham formula $\left(B_{c2}(\theta)\sin\theta/B_{c2}^{\parallel}\right)^2 + \left|\frac{B_{c2}(\theta)\cos\theta}{B_{c2}^{\perp}}\right| = 1$ and the 3D anisotropic mass model $B_{c2}(\theta,T) = \gamma B_{c2}^{\perp}(T)(sin^2\theta + \gamma^2 cos^2\theta)^{-\frac{1}{2}}$ with $\gamma = B_{c2}^{\parallel}/B_{c2}^{\perp}$, respectively. (c) Temperature dependent square resistance of 5.0-nm-thick TaSe$_2$ device ($T_c^{onset}$~2.02 K, $T_c^{zero}$~1.72 K) for $T > 0.5$ K at different out-of-plane magnetic fields. For two curves in same color, the bottom ones show results with filters and the top ones show results without filters. (d) $R_S$ versus $T^{-1}$ curves corresponding to (d).

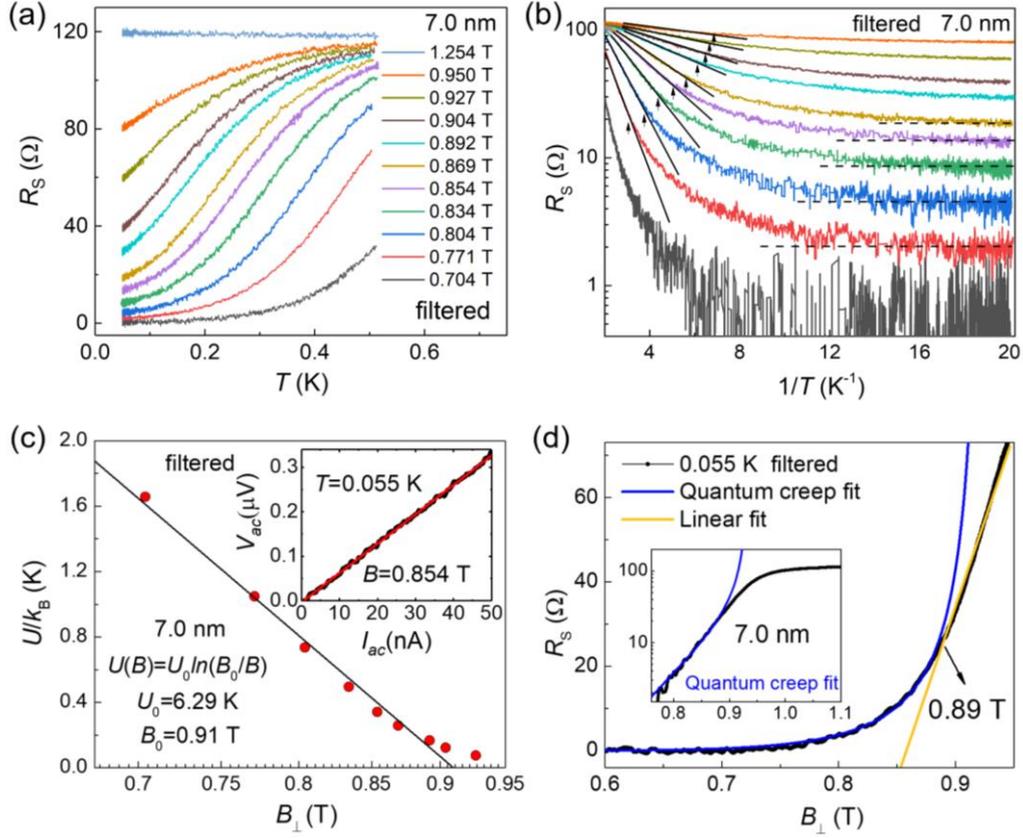

**Figure 3.** Intrinsic anomalous metallic state. (a) Filtered $R_S(T)$ curves of 7.0-nm-thick TaSe$_2$ ($T_c^{onset}$ ~2.05 K, $T_c^{zero}$ ~1.90 K) for $T$ < 0.5 K, with magnetic fields ranging from 0.704 T to 1.254 T. (b) $R_S$ versus $T^{-1}$ curves corresponding to (a). The black solid lines indicate thermally activated behavior. The arrows separate the thermally activated state at the higher temperatures and the saturated state at lower temperatures. The black dashed lines are eye-guides for marking the $R$ saturations. (c) Log $B_\perp$-dependent activation energy $U$, obtained from the slope of the linear portion in (b). Inset: $VI$ curve at 0.055 K and 0.854 T. The red line is a linear fitting curve. (d) $R_S(B_\perp)$ curve of 7.0-nm-thick TaSe$_2$ at 0.055 K, can be well fitted by quantum creep theory (Eq.(1)) in the relatively low magnetic field region (the solid blue line). The $R_S(B_\perp)$ curve exhibits a linear magnetic field dependence above 0.89 T. The inset shows the same data from 0.75 T to 1.1 T, with $R_S$ plotted in log scale.

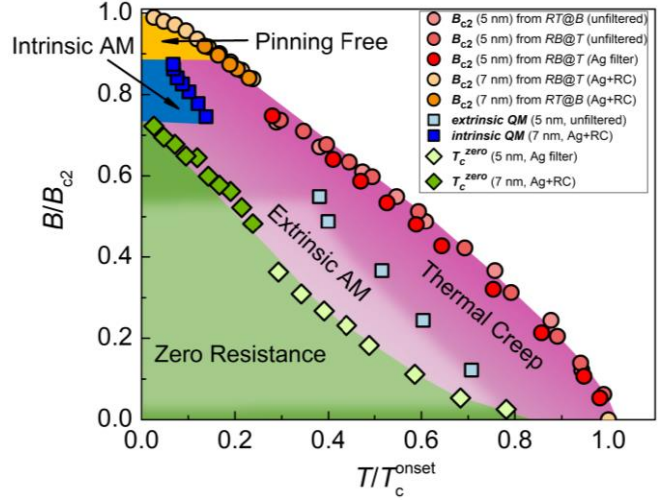

**Figure 4.** Phase diagram of 2D TaSe$_2$. The temperature and perpendicular magnetic field are normalized by $T_c^{onset}$ and $B_{c2}$, respectively. The blue region and white transparent region represent intrinsic and extrinsic anomalous metallic (AM) states, respectively. The boundary between the thermally activated vortex creep region and the intrinsic (extrinsic) AM region is determined from the thermal activated analysis in Fig. 3(b) (Supporting Information Fig. S3).

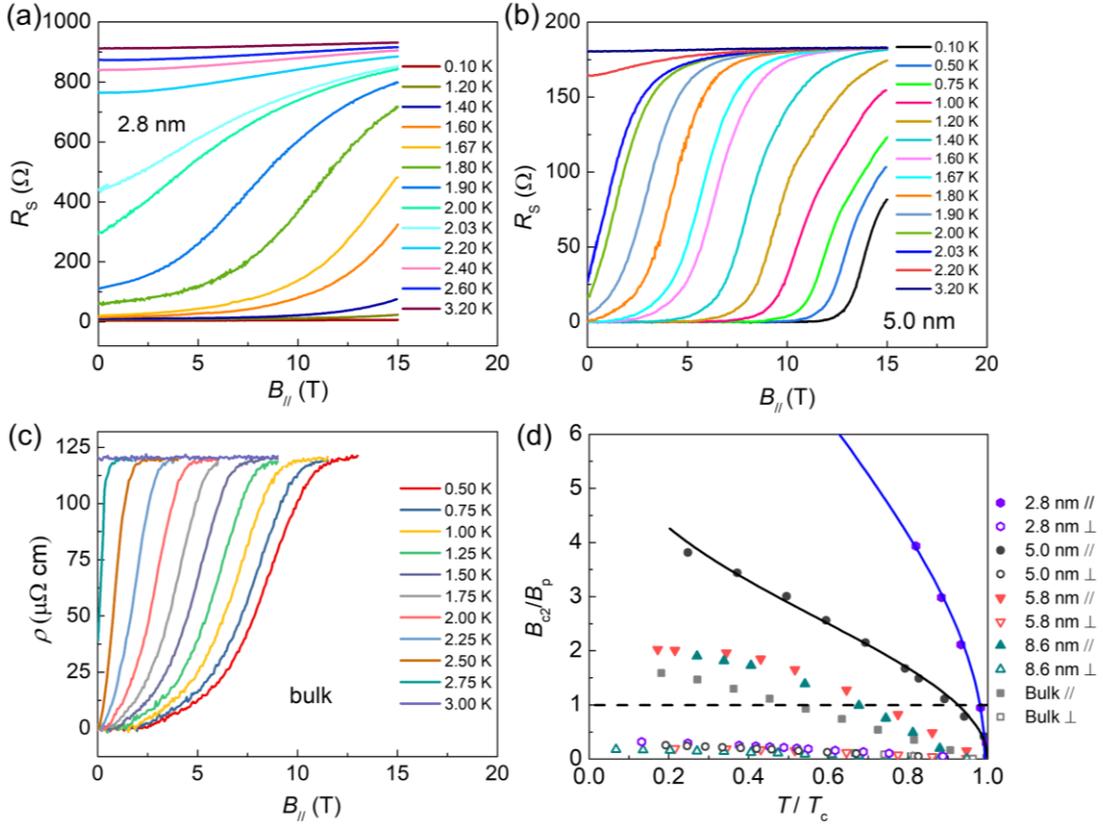

**Figure 5.** Ising superconductivity in 4$Ha$-TaSe$_2$ nanodevices. The in-plane magnetic field dependence

of the square resistance in 2.8 nm (a) and 5.0 nm (b) thick TaSe$_2$ devices at different temperatures from 0.10 K to 3.20 K. (c) Magnetoresistivity of bulk TaSe$_2$ as a function of in-plane magnetic field at different temperatures. (d) The critical field $B_{c2}/B_p$ as a function of transition temperature $T/T_c$ for TaSe$_2$ samples of different thickness under both out-of-plane $B_\perp$ (open symbols) and in-plane $B_{//}$ (filled symbols) magnetic fields. The blue and black lines are the fitting curves of the Zeeman-protected Ising superconductivity model (Eq. (2)) for 2.8 nm and 5.0 nm devices. For bulk TaSe$_2$, the $B_{c2//}$ exceeds its $B_p$ when $T/T_c < 0.5$. The black dashed line marks the $B_p$.

## ASSOCIATED CONTENT

**Supporting Information**

The Supporting Information is available free of charge at https://pubs.acs.org.

Crystal growth method, device fabrication method, sample characterization information, transport measurements details, schematics of the measurement circuits, transport data of the bulk 4$Ha$-TaSe$_2$, $VI$ curve and d$V$/d$I$ curve of the 7.0-nm-thick TaSe$_2$ device, intrinsic anomalous metallic state in 4.4-nm-thick 4$Ha$-TaSe$_2$ device, magnetoresistance of a 2.8-nm-thick, 5.0-nm-thick, 5.8-nm-thick and 8.6-nm-thick TaSe$_2$ devices under out-of-plane and in-plane magnetic fields, Bose metal fitting of the anomalous metallic state for TaSe$_2$ device, and Hall results of 5.0 nm 4$Ha$-TaSe$_2$ nanodevice and bulk (PDF)

## AUTHOR INFORMATION

**Corresponding Authors**

*jianwangphysics@pku.edu.cn (J.W.)

*xilin@pku.edu.cn (X.L.)

*yiliu@ruc.edu.cn (Yi L.)**Author contributions:** J.W. designed and instructed the research. P.Y., Y.J.L. synthesized and characterized bulk single crystals; P.Y., J.G., Z.Y. and Z.W. fabricated the devices; Y.X., P.Y. did the transport measurements with the assistance of Y.Z.L., C.X., M.T.; J.Y., X.L., J.L., H.J. helped with the ultralow temperature measurements in dilution refrigerators; F.Y. and P.Y. did the TEM characterization; P.Q. carried out the Raman spectra; Yi L. contributed to the theoretical interpretation. Y.X., P.Y. and J.W.

14 / 30

analyzed the data and wrote the manuscript. All authors discussed the results and commented on the manuscript.

‡ These authors contributed equally to this work.

**Notes**

The authors declare no conflict of interest.

ACKNOWLEDGMENTS.

We thank Tianheng Wei, Pinyuan Wang, Qingzheng Qiu, Liqin Huang, Shichao Qi, Wenlu Lin, Wei Han for help in filters fabrication and test, Shuai Yuan for help in ultralow temperature transport measurements, and Cong Wang for help in TEM sample fabrication. **Funding:** This work is financially supported by the National Natural Science Foundation of China (11888101), the National Key R&D Program of China (2018YFA0305604, 2017YFA0303302), the National Natural Science Foundation of China (11974430, 11774008,11704414, 12174442), Beijing Natural Science Foundation (Z180010), the Strategic Priority Research Program of Chinese Academy of Sciences (XDB28000000).

# Supporting Information for:

# Extrinsic and Intrinsic Anomalous Metallic States in Transition Metal Dichalcogenide Ising Superconductors

Ying Xing,$^{//,†,‡}$ Pu Yang,$^{//,\Delta,‡}$ Jun Ge,$^{†,‡}$ Jiaojie Yan,$^{†,‡}$ Jiawei Luo,$^{†}$ Haoran Ji,$^{†}$ Zeyan Yang,$^{//}$ Yongjie Li,$^{//}$ Zijia Wang,$^{//}$ Yanzhao Liu,$^{†}$ Feng Yang,$^{//}$ Ping Qiu,$^{//}$ Chuanying Xi,$^{◊}$ Mingliang Tian,$^{◊}$ Yi Liu,$^{#,*}$ Xi Lin,$^{†,§,⊥,*}$ Jian Wang,$^{†,§,⊥,*}$

$^{//}$State Key Laboratory of Heavy Oil Processing, College of New Energy and Materials, China University of Petroleum, Beijing 102249, China.

$^{†}$International Center for Quantum Materials, School of Physics, Peking University, Beijing 100871, China.

$^{\Delta}$College of Chemistry, Beijing Normal University, Beijing 100875, China.

$^{◊}$High Magnetic Field Laboratory, Chinese Academy of Sciences, Hefei 230031, China.

$^{#}$Department of Physics, Renmin University of China, Beijing 100872, China.

$^{§}$CAS Center for Excellence in Topological Quantum Computation, University of Chinese Academy of Sciences, Beijing 100190, China.

$^{⊥}$Beijing Academy of Quantum Information Sciences, Beijing 100193, China.

‡ These authors contributed equally to this work.

* Corresponding author. E-mail: jianwangphysics@pku.edu.cn (J.W.); xilin@pku.edu.cn (X.L.); yiliu@ruc.edu.cn (Yi L.)

**This PDF file includes:**

Materials and Methods

Figures (S1-S11)

References (1-3)

## MATERIALS AND METHODS

**Crystal growth.**

4$Ha$-TaSe$_2$ single crystals were prepared by the iodine-assisted vapor transport method. First, Ta (99.99%) and Se (99.999%) powder were mixed with the ratio of 1:2, ground adequately, sealed into an evacuated quartz ampoule, heated to 750 °C and kept for 3 days. Subsequently, the mixture polycrystalline TaSe$_2$ was obtained and reground with iodine (8 mg/cm$^3$), then sealed into evacuated quartz tubes, reacted for 7 days in a two-zone furnace where the temperature was set as 750 °C and 690 °C. The temperature gradient of 3 K/cm was applied along the evacuated quartz tube inside a two-zone furnace. Finally, shiny, flaky single crystals with a typical size of about 3 × 3 × 0.05 mm$^3$ grew up at the colder side of the tube.

**Devices fabrication.**

Few-layer flakes were mechanically exfoliated from high quality bulk single crystals by Scotch tapes. The ultrathin samples (2.8 to 52.0 nm) were first transferred onto SiO$_2$/Si substrates and then identified by optical microscopy and AFM. Standard four-probe or Hall structure configuration devices were fabricated using e-beam lithography (FEI Helios NanoLab 600i DualBeam System). The Pd/Au (6.5/80 nm) electrodes were deposited using an electron-beam evaporator (LJUHV E-400L) followed by lift-off in acetone. To obtain a clean interface between the electrodes and the sample, *in situ* argon plasma was employed to remove the resist residues before metal evaporation without breaking the vacuum. No additional protective layer is used during the whole process.

**Sample characterization.**

The sample characterization data were collected from Optical Microscope (Olympus BX51), XRD (Bruker AXS, using Cu Kα radiation in the 2$\theta$ range of 10°~90°), EDS (Oxford X-MaxN 80),

Raman spectroscopy (Renishaw Centrus 0JPN82, excitation wavelength~532 nm), TEM (JEM 2100; acceleration voltage, 200 kV), SEM (FEI Helios NanoLab 600i DualBeam System, 2 kV for SEM), AFM (Dimension Icon, Bruker).

**Transport measurements.**

The Ising superconductivity measurements were carried out with a standard four-probe geometry in a Physical Property Measurement System (Quantum Design, PPMS-16, d.c. technique), with the Helium-3 option for temperature down to 0.5 K and dilution option down to 0.05 K. The contacts on the single crystals were made by applying the sliver paste on the top surface (001) of 4$Ha$-TaSe$_2$, with contact resistance of several Ohms. The contact resistance between 4$Ha$-TaSe$_2$ flake and Pd/Au electrodes is 10 ~ 300 Ω. The excitation current is ~ 1 mA for bulk, and 10 nA~500 nA for devices, respectively. Angular dependence of the magnetoresistance was measured in a Helium-3 cryostat by using a rotation holder. The experiments on verifying anomalous metal were carried out on Leiden CF450 ($T$ > 0.5 K) and Leiden CF-CS81-600 ($T$ < 0.5 K). The rate for magnetic field sweeping was 1 mT/s, and the sweeping rates of temperature changed with different temperatures. In CF-CS81-600 system, for $T$ >200 mK, it was about 0.3 mK/s, and for $T$ <100 mK, the rate was much slower and it took about 20 minutes to cool down from 100 mK to 55 mK. In CF-CS81-600 system, home-made room temperature RC filters (3rd-order RC filters with resistors of 510 Ω, 820 Ω, 1500 Ω and capacitors of 4.7 nF, 2.2 nF, 1.1 nF for each stage respectively) and home-made silver-epoxy filters (installed at a plate with the same temperature as the sample) are connected in series with each lead of samples. The combination of two kind of filters reaches the analyzer's noise floor for frequency above 600 MHz and attenuation is more than 45 dB for frequency larger than 300 kHz.[1] Moreover, in a prior work, the temperature of high quality GaAs/AlGaAs sample in the same instrument (Leiden CF-CS81-600 DR) with same filters is cooled down to 25 mK and demonstrated to be the same as the electron temperature according to the activation behaviors of fractional quantum Hall states. [1] The controlled experiments were carried out in Leiden CF450 system, with and without the home-made room temperature silver-epoxy filters.

**FIGURES**

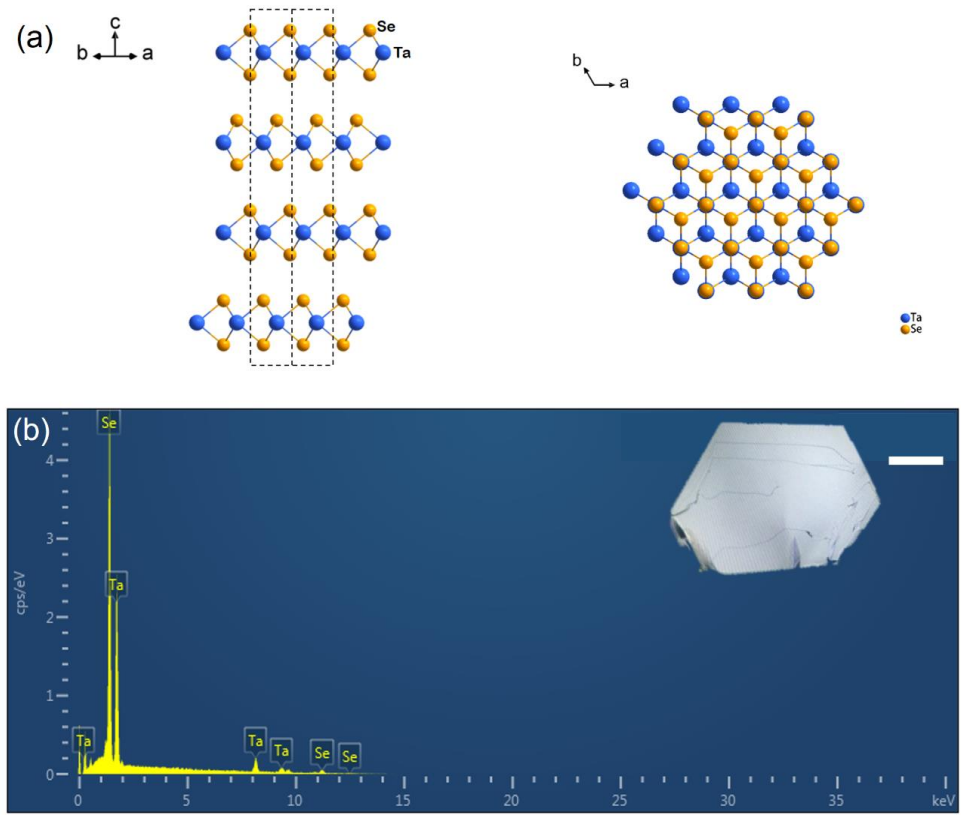

**Fig. S1.** Sample characterizations of 4$Ha$-TaSe$_2$. (a) Schematic crystal structure (left: side view, right: top view) of 4$Ha$-TaSe$_2$, where the Ta and Se atoms are colored in blue and yellow, respectively. The in-plane inversion symmetry is broken in each individual layer, and global inversion symmetry is also broken in bulk after stacking. (b) The energy dispersive spectrometer pattern of the 4$Ha$-TaSe$_2$ single crystal. The ratio of Ta to Se is determined to be 1.1:2.0, which is an average result with different areas in the same crystal. Inset: optical image of a typical bulk 4$Ha$-TaSe$_2$, the scale bar corresponds to 1 mm.

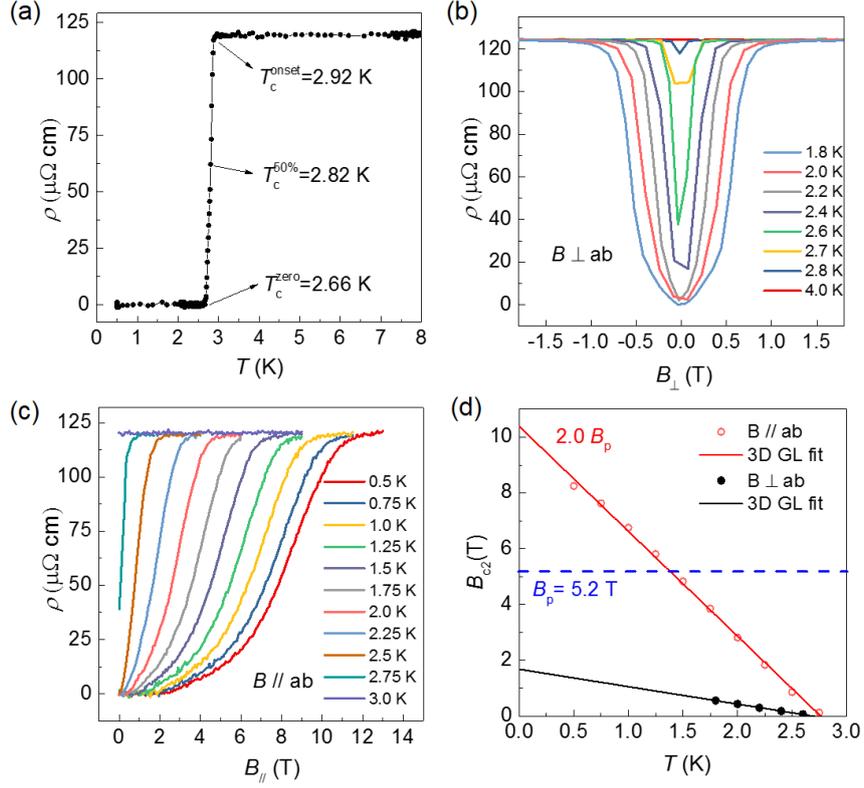

**Fig. S2.** Transport data of the bulk 4*Ha*-TaSe$_2$. (a) The temperature dependence of the resistivity in bulk 4*Ha*-TaSe$_2$ single crystal ranging from 0.5 K to 8.0 K. The onset ($T_c^{onset}$), middle ($T_c^{50\%}$, as defined by the temperature where the resistivity reaches 50% of the normal state value) and zero-resistance ($T_c^{zero}$) superconducting critical transition temperatures are 2.92 K, 2.82 K, 2.66 K respectively. (b) (c) Resistivity as a function of magnetic field applied perpendicular and parallel to the surface of the sample at different temperatures, respectively. (d) The critical field $B_{c2}$ (50% normal resistivity) in both directions as a function of temperature *T*. The solid lines are fitting curves by 3D Ginzburg-Landau (GL) theory[2] $B_{C2} \propto (T_c - T)$. The anisotropic parameter $\gamma = B_{c2//}/B_{c2\perp}$ is about 6 for bulk TaSe$_2$.

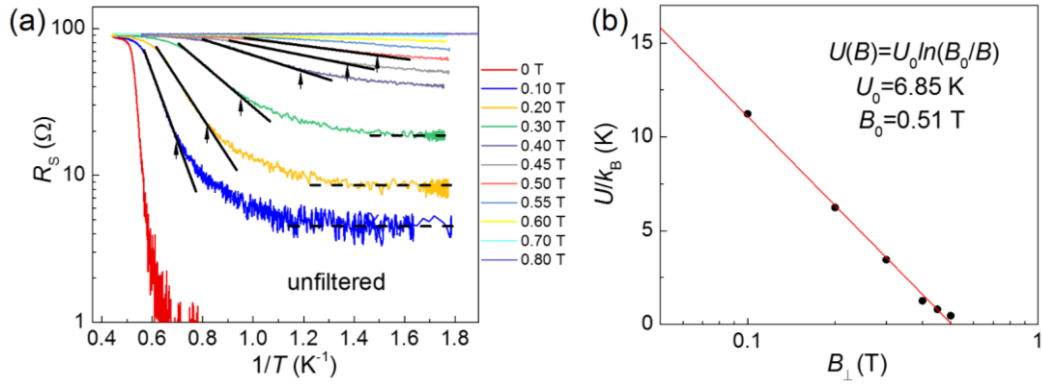

**Fig. S3.** Thermal activated analysis of unfiltered data in the main text of Fig. 2(d). (a) log $R_S$ vs. $1/T$ at various magnetic fields. The solid lines are fitting results from the Arrhenius relation. The dashed lines are guiding lines for $R$ saturation. (b) Field dependence of $U/k_B$. The red solid line is a linear fitting.

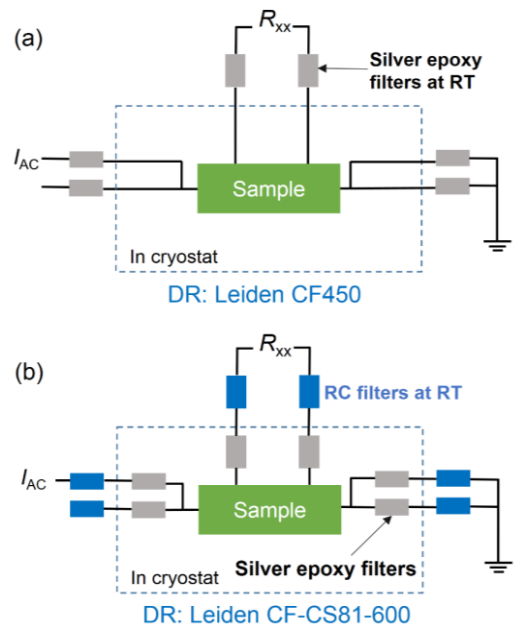

**Fig. S4.** Schematics of the measurement circuits. (a) Schematics of the measurement circuit in Leiden CF450 system for $T > 0.5$ K. The dashed blue square represents the cryogenic system. The filtered data shown in Fig. 2(c) in the main text were measured by home-made silver-epoxy filters installed at room temperature. The unfiltered data were measured by removing the silver-epoxy filters. (b) Schematics of the measurement circuit in Leiden CF-CS81-600 for $T < 0.5$ K. The filtered

data shown in Fig. 3 in the main text were measured by connecting home-made silver-epoxy filters (cut-off frequency around 470 kHz) at low temperature stage and resistor-capacitor (RC) filters (cut-off frequency designed for 22 kHz) at room temperature, in series with each lead of sample. The performance of the individual home-made filters and their combination can be found in the work of Wang *et al* [1]. The combination of the silver-epoxy filter and RC filter in series behaves an attenuation of more than 90 dB above 1GHz, reaching the noise floor (the least level of noise in this experimental setup). The attenuation in the MHz range reduces to ~0.0001% of the initial noise power.

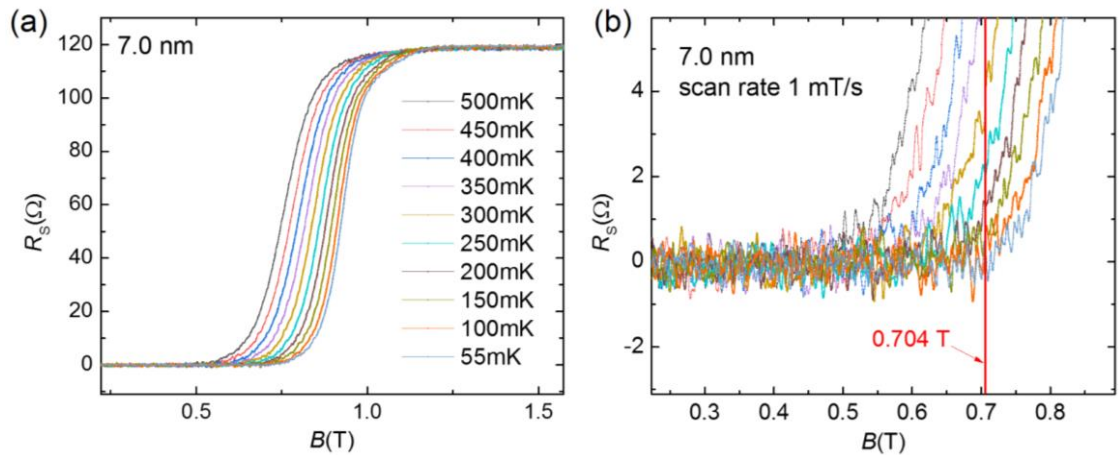

**Fig. S5.** (a) $R_s(B)$ curves of the 7.0 nm $4Ha$-TaSe$_2$ nanodevice, the resistance just reaches zero at 55 mK at 0.704 T. (b) Close-up of the same data near the zero resistance in (a). The noise level is around 0.8 Ω.

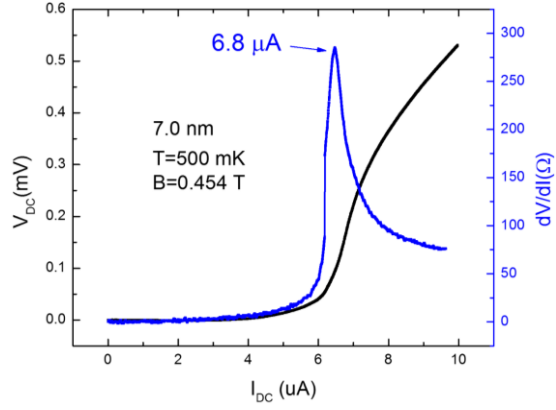

**Fig. S6.** *VI* curve (black line) and d*V*/d*I* curve (blue line) of the 7.0-nm-thick TaSe$_2$ device (the device shown in Fig.3 of the main text) measured at 500 mK and 0.454 T, with superconducting critical current ~ 6.8 μA.

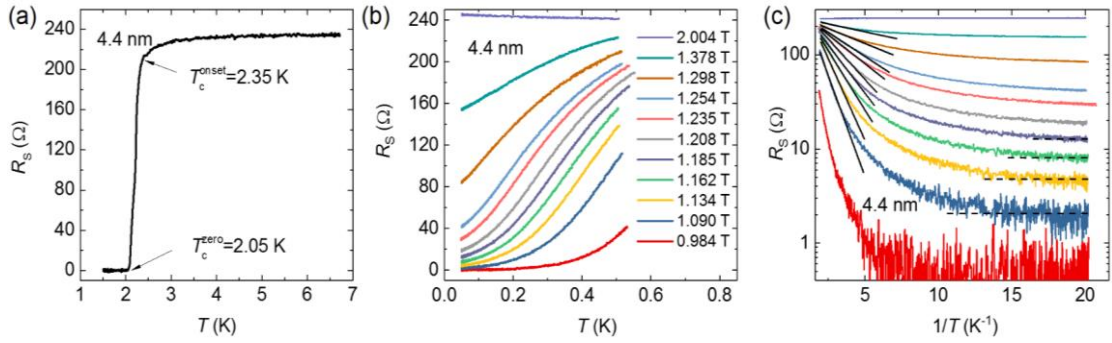

**Fig. S7.** Intrinsic anomalous metallic state in 4.4-nm-thick 4*Ha*-TaSe$_2$ device measured in Leiden CF-CS81-600 system with low-temperature silver epoxy filters combined with room-temperature RC filters. (a) $R_S(T)$ curve from 1.5 K to 6.5 K ($T_c^{onset}$~2.35 K, $T_c^{zero}$~2.05 K) at zero magnetic field. (b) Filtered $R_S(T)$ curves under different magnetic fields ranging from 0.984 T to 2.004 T. (c) $R_S$ versus $T^{-1}$ corresponding to (b). The black solid lines indicate thermally activated behavior. The black dashed lines are eye-guides for marking the $R_S$ saturations. The 4.4-nm-thick 4*Ha*-TaSe$_2$ device shows consistent results with 7.0 nm device. Therefore, the stacking order in 4*Ha*-TaSe$_2$ makes little difference on the physical properties studied in this work.

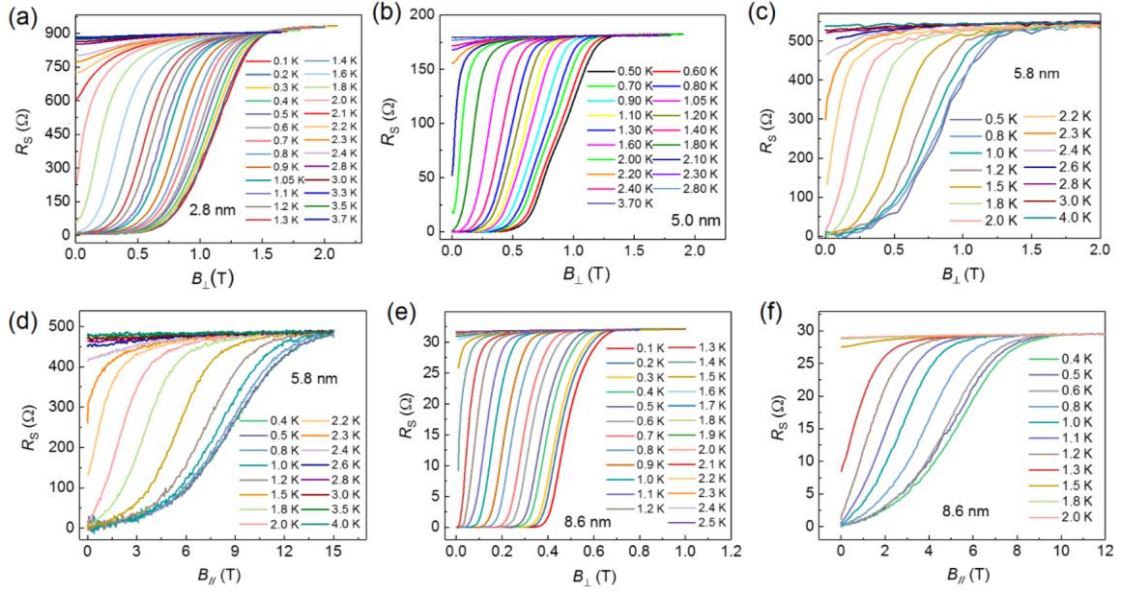

**Fig. S8.** Magnetoresistance of a 2.8-nm-thick (a), 5.0-nm-thick (b), 5.8-nm-thick (c)(d) and 8.6-nm-thick (e)(f) TaSe$_2$ devices under out-of-plane and in-plane magnetic fields.

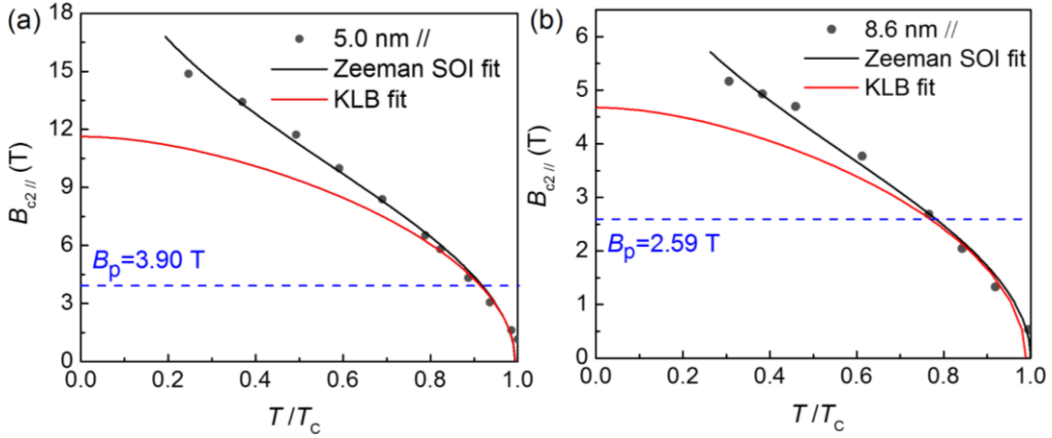

**Fig. S9.** The in-plane critical field $B_{c2//}$ as a function of transition temperature $T/T_c$ for 5.0 nm (a) and 8.6 nm (b) 4$Ha$-TaSe$_2$ devices. The red and black curves are the fitting curves based on the microscopic Klemm-Luther-Beasley (KLB) theory[3] and the Zeeman-type spin-orbit interaction, respectively. Our data agree better with Zeeman-type spin-orbit interaction expression. The KLB theory can fit well near $T_c$, but deviates obviously at $T/T_c < 0.6$.

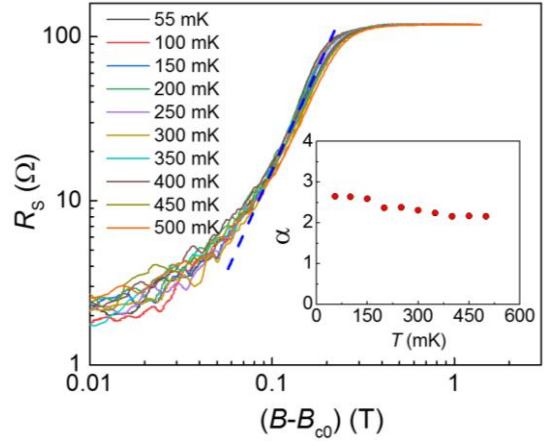

**Fig. S10.** An alternative explanation of the anomalous metallic state for TaSe$_2$ device is the Bose metal theory, which describes a gapless, non-superfluid state in the zero temperature limit. In the framework of this theory, the small residual resistance at finite magnetic fields results from uncondensed Cooper pairs and vortices, and the square resistance of the metallic state is proportional to $(B - B_{c0})^{2\nu}$, where $B_{c0}$ is the critical field of superconductor to Bose metal transition and $\nu$ is the exponent of the superfluid correlation length. Here, $B_{c0}$ is defined as the critical field of zero resistance within the measurement resolution. The magnetoresistance $R_S(B - B_{c0})$ curves of 7.0-nm-thick TaSe$_2$ device is plotted in the double logarithmic scale at various temperatures from 55 mK to 500 mK with out-of-plane magnetic fields. The dashed blue line indicates the slope of the $R_S(B - B_{c0})$ curves in the low temperature regime. The isotherms below 500 mK collapse to a single curve with the slopes ($\alpha = 2\nu$) between 3.0 and 3.5. The inset shows the slope of the magnetoresistance as a function of temperature, which yields a critical exponent $\nu \sim 1.69$ at ultralow temperatures.

29 / 30

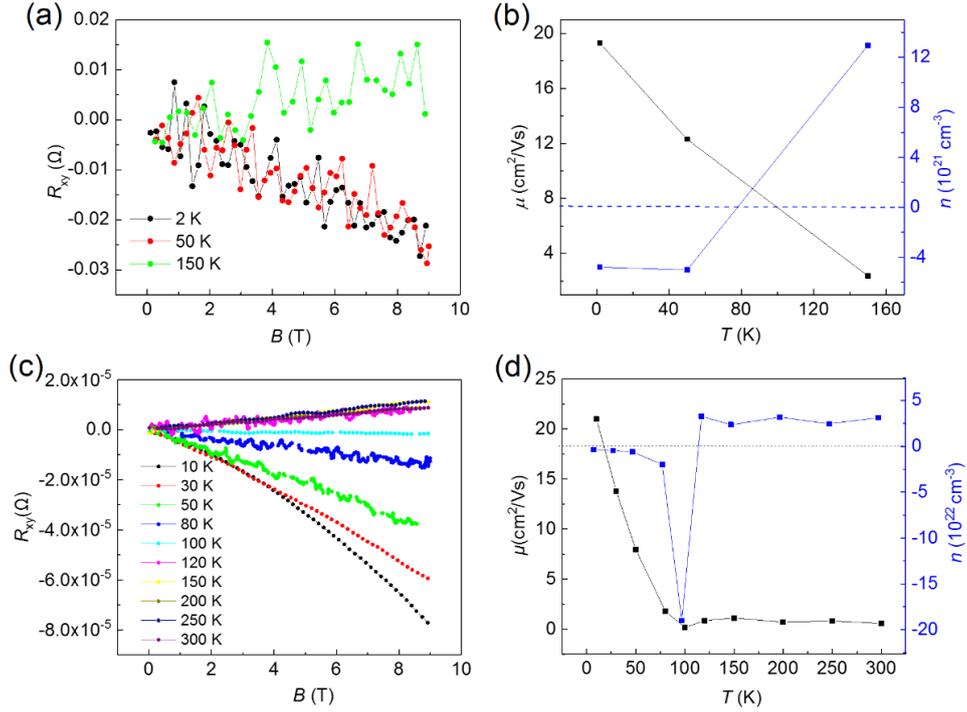

**Fig. S11.** Hall results of a 5.0 nm 4$Ha$-TaSe$_2$ nanodevice and bulk. $R_{xy}$ vs magnetic field curves at different temperatures of TaSe$_2$ nanodevice (a) and bulk (c). The carrier mobility and carrier density of TaSe$_2$ nanodevice (b) and bulk (d) at different temperatures obtained from data in (a) & (c). The electron mean free path $l$ is estimated to 38.6 nm, larger than in-plane $\xi_{GL}$ (~16.5 nm in 5.0 nm TaSe$_2$). This is consistent with clean superconductors, in which the Bardeen-Cooper-Schrieffer coherence length $\xi_0 = 1.35\xi_{GL} < l$. While in dirty superconductors, $\xi_0$ is much larger than $l$, [2] not applicable to our result.